\documentclass{article}
\usepackage[utf8]{inputenc}
\usepackage{authblk}
\usepackage[dvips]{graphicx}
\graphicspath{{noiseimages/}}
\usepackage{xcolor}
\usepackage[T2A]{fontenc}
\usepackage{tikz}
\usepackage{pgfplots}
\pgfplotsset{compat=1.7}
\usetikzlibrary{intersections, pgfplots.fillbetween}
\usetikzlibrary{decorations.pathmorphing}
\usepackage[mode=buildnew]{standalone}
\usepackage[toc,page]{appendix}
\usepackage{amsmath}
\usepackage{amssymb}
\usepackage{float}
\usepackage{comment}
\usepackage{a4wide}
\usepackage[english]{babel}
\usepackage{hyperref}
\definecolor{urlcolor}{HTML}{990000}
\definecolor{linkcolor}{HTML}{005F5F}
\hypersetup{pdfstartview=FitH,  linkcolor=linkcolor,urlcolor=urlcolor, colorlinks=true,citecolor=blue}
\usepackage[page]{appendix}

\newcommand{\ind}{{-\frac12+i\mu}}

\newcommand{\w}{\omega}

\renewcommand{\hat}{}

\author[1,2]{E.T.Akhmedov}
\author[1,2]{P.A.Anempodistov}
\author[1,2]{K.V.Bazarov}
\author[1,2]{D.V.Diakonov}
\author[3,4]{U.Moschella}
\affil[1]{Moscow Institute of Physics and Technology, Institutskii per. 9, 141700, Dolgoprudny, Russia}
\affil[2]{ Institute for Theoretical and Experimental Physics, B. Cheremushkinskaya 25, 117218, Moscow, Russia}
\affil[3]{Universit\`a degli Studi dell'Insubria - Dipartimento DiSAT, Via Valleggio 11 - 22100 Como - Italy}
\affil[4]{INFN, Sez di Milano, Via Celoria 16, 20146, Milano - Italy}

\title{\textcolor{black}{Heating up an environment around black holes and inside de Sitter space}}
\begin{document}

\numberwithin{equation}{section}

\maketitle

\begin{abstract}

We study quantum fields on spacetimes having a bifurcate Killing horizon by allowing the possibility that left-- and right-- (in--going and out--going) modes have different temperatures. 
We consider in particular the Rindler for both massless and massive fields, the static de Sitter and Schwarzschild black hole backgrounds for massive fields.
We find that in all three cases, when any of the temperatures is different from the canonical one (Unruh, Hawking and Gibbons--Hawking, correspondingly) the correlation functions have extra singularities at the horizon. 

\end{abstract}
\newpage

\tableofcontents

\newpage

\section{Introduction }

There is a well-known relation between the existence of a bifurcate Killing horizon on certain spacetime manifolds and the temperature of some specially privileged equilibrium thermal states.
The Rindler, the  de Sitter, and the Schwarzschild spacetimes  are examples of such situation \cite{kaywald}.

While there is a vast literature devoted to the study of the above states, little attention has been paid to other possible thermal or pseudo-thermal states, let alone to what happens when interactions are switched on:  does the fluctuation--dissipation theorem still work as in flat space? Are the privileged  states attractor solutions of some sort of kinetic equations with exact modes instead of plane waves?  Does thermalization of a given initial state happens in {\it strongly} curved space--times without substantial backreaction on the gravitational background? The backreaction during thermalization seems to be negligible practically for any reasonable (Hadamard) initial state in flat space--time \cite{LL10,Kamenev}.

When attempting to address some of the above questions  one encounters many underwater stones; for example, describing thermal states other than the Gibbons-Hawking one \cite{HG} in the static de Sitter--Rindler wedge \cite{deSitter,Akhmedov:2020qxd} is already non trivial. Also what would be a thermal state does not necessarily posses all the same properties as a regular thermal state in flat space \cite{Popov:2017xut}. Furthermore,
if time translations are not a symmetry,  as is the case  the the global Lanczos' spherical coordinate system \cite{lanczos}, secular divergences arise both in distributions and in anomalous averages  \cite{Krotov:2010ma,Akhmedov:2013vka,Akhmedov:2012dn,Akhmedov:2019cfd}. The latter can be resummed for fields whose mass is bigger than a critical mass, via analogues of kinetic equations for both the distributions and anomalous averages. Such kinetic equations do not have Planckian or Boltzmannian solutions. Moreover they have exploding solutions. Such a situation when thermalization does not happen without taking into account backreaction on the background field is similar to the one encountered in constant electric field \cite{Akhmedov:2014hfa,Akhmedov:2014doa}.

As regards the Schwarzschild geometry and the black hole radiation, there are three distinguished states which are usually considered: the Boulware \cite{Boulware:1974dm},  Unruh \cite{Unruh:1976db} and Hartle-Hawking states.   \cite{Hartle:1976tp,Candelas:1980zt}.
The Boulware state is the vacuum of both the in--going and out--going modes of the Schwarzschild background; the  Unruh state is the vacuum of the in--going modes and has the Planckian distribution at the Hawking temperature  for the out--going modes; finally, in the Hartle--Hawking state  both the in--going and the out--going modes are thermally distributed at the Hawking temperature.

Can any of the aforementioned quantum states actually  describe the fate the quantum field 
at the end of the collapse \cite{Hawking:1974sw}?
Actually, in the formation of a black hole  process one has to consider a different basis of modes \cite{Akhmedov:2015xwa} where the in--going and out--going modes are not treated as independent but rather a linear combination of them which is regular at the center of the collapsing star. The corresponding state is inequivalent to either the Boulware, the Unruh or the Hartle--Hawking state.
The question then arises whether the initial state before the collapse can thermalise to any of the above mentioned states.

A second set of questions regards the behaviour of black holes surrounded by a gas with a temperature different from the Hawking one.
Can one heat up a black hole by surrounding it with a gas of temperature different from the Hawking one? And how the heating works in detail?
In concrete astrophysical situations black holes indeed are surrounded by accretion disks which definitely have nothing to do with the Hawking radiation. It goes without saying about primordial black holes in early universe.


In this paper even without performing loop calculations (where the heating process is actually seen) we will argue that the answer to the first question in the last paragraph seems to be negative. Both in the static de Sitter space (see also \cite{Akhmedov:2020qxd}) and in the black hole background all correlation functions with temperatures different from the Hawking one have anomalous singularities at the horizon.

To simplify our discussion for the beginning we consider real scalar field theory in two dimensions, in
either the Rindler, the static de Sitter or two--dimensional analog of the Schwarzschild background. We study the properties of the tree--level Wightman functions for a class of time translation invariant states, which include states having different temperatures for the ingoing and the outgoing modes. 
In all cases when the temperatures are different from the Unruh, Gibbons--Hawking and Hawking ones in the corresponding situations there are anomalous singularities of the correlation functions at the horizon.

\section{Rindler space--time}

To put the results of the paper in perspective we start by discussing the Rindler spacetime. This section mainly  contains a recapitulation of known facts. However some of them are new.
\subsection{Geometry, modes and Wightman function}

\label{rindler}

The coordinate system for the Rindler right wedge of the Minkowski spacetime is obtained by applying the one-parameter subgroup of boosts, which leaves invariant the wedge, to points of, say,  the half-line $t = 0$, $x = e^{\alpha\xi}/\alpha>0$:
%
\begin{eqnarray}
\label{coordinates1}
X(\eta,\xi)=\left(
\begin{array}{c}
 t \\
 x \\
\end{array}
\right) =     \left(
\begin{array}{cc}
 \cosh (\alpha  \eta ) & \sinh (\alpha  \eta ) \\
 \sinh (\alpha  \eta ) & \cosh (\alpha  \eta ) \\
\end{array}
\right) \left(
\begin{array}{c}
 0 \\
 \frac{1 }{\alpha } e^{\alpha  \xi}\\
\end{array}
\right) =\left(
\begin{array}{c}
\frac 1 \alpha e^{\alpha  \xi } \sinh (\alpha  \eta ) \\
\frac 1 \alpha e^{\alpha  \xi } \cosh (\alpha  \eta ) \\
\end{array}
\right) .
\end{eqnarray}
Here $\eta$ is the parameter of the  subgroup and is interpreted as the Rindler time, $\xi$ is the space coordinate and $\alpha$  is the proper acceleration. For real values of $\eta$ and $\xi$ the  Rindler coordinates \eqref{coordinates1} 
cover only the right wedge; this is causally disconnected from the left wedge. The half-lines $x=\pm t$ with $x>0$ are  the past and the future horizons.
When  $\eta$ and $\xi$ are complex  
they cover the full  Minkowski spacetime;
in the following  we will set the acceleration to one ($\alpha=1$).

In the above coordinates the metric is static and conformal to Minkowski
\begin{align}
\label{metricRindler}
    ds^2=e^{2 \xi}\Big(d\eta^2-d\xi^2\Big)\, ;
\end{align}
the invariant interval between two events in the wedge is given by
\begin{align}
\label{geodistRindler}
   L_{12}= (X_1-X_2)^2= 2e^{(\xi_1+\xi_2)}\cosh(\eta_2-\eta_1)
-e^{2\xi_2}-e^{2\xi_1}.
    \end{align}
For the future reference please note the obvious symmetry of the interval in the exchange

\begin{equation}
    \xi_1\longleftrightarrow \xi_2.
    \label{symmetry}
    \end{equation}
Lorentz transformations of the wedge corresponds to  (time) translations in the $\eta$ variable:
 $ \eta\to\eta +\gamma$;
dilatations in Minkowski space $X\to e^{\beta}X$ correspond to the the shift
   $ \xi \to \xi + \beta$.
%
\begin{figure}\centering\includestandalone[width=0.6\textwidth]{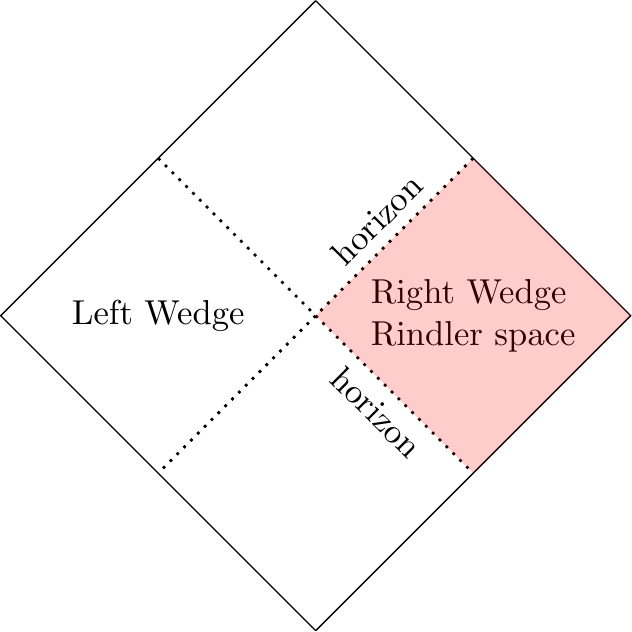}\caption{Penrose diagram of the Rindler space. The Rindler space is bordered by a Killing horizon.}\label{dssp}\end{figure}
 As regards the light cone variables
 \begin{eqnarray}
\label{coordinatesuv}
u =  t-x  = - e^{ (\xi -\eta)}  = -  e^{- U  },
 \ \ \ \ \ v =  t+x =   e^{  (\xi +\eta)}  = e^{  V},
\label{lorentz0}
\end{eqnarray}
they are transformed as follows:
 \label{coordinatesuv}
 \begin{eqnarray}
 \begin{array}{l}
u \to e^{- \gamma   }  u,\\ \\
v \to  e^{  \gamma} v ,
\end{array} \ \  \begin{array}{l}
u \to e^{\beta   }  u,\\ \\
v \to  e^{  \beta} v,
\end{array}\  \
\begin{array}{l}
U \to U+\gamma,\\ \\
V \to  V+\gamma,
\end{array}\  \
 \begin{array}{l}
U \to U-\beta,\\ \\
V \to  V+ \beta.
\end{array}
\label{lorentz0}
\end{eqnarray}
Though geodetically incomplete, the Rindler wedge is a globally hyperbolic manifold in itself; of course a Cauchy surface,  
say $\eta=0$, 
is not a Cauchy surface for the  whole Minkowski spacetime. As a consequence the  modes constructed by  canonical quantization  in the Rindler wedge  do not constitute a basis for the  whole Minkowski space--time. It is well-known that to obtain general Hilbert space representations of the fields, including the ones carrying unitary representations of the Poincar\'e group, one needs to construct also the modes defined in the left wedge  \cite{Unruh:1983ms}. A less known but powerful alternative is to resort to the theory of generalized Bogoliubov transformations\footnote{Starting from pure states generalized Bogoliubov transformations may produce mixed states  while standard Bogoliubov transformations cannot.} which  makes use only of the modes of the right wedge \cite{ms,ms2}.
The Klein--Gordon equation for a massive scalar field in two-dimensions 
is as follows:

\begin{align}
\label{eqmassive}
\Big(\partial_\eta^2-\partial_\xi^2+e^{2\xi}m^2\Big)\varphi(\eta,\xi)=0.
\end{align}
By separating the variables one gets a Schrodinger eigenvalue (textbook) problem in an exponential potential $V(\xi)=m^2 e^{2\xi}$.
Normalizable modes 
are proportional to Macdonald functions $K_{i\omega}(me^\xi)$ which are linear combinations of left-moving and right-moving waves.

The canonical field operator may  be expanded as follows
\begin{align}
\label{operatormassiveRindler}
    \hat{\varphi}(X) =\frac 1 {\pi}\int_{0}^{+\infty}  \, \bigg(e^{-i\omega \eta}\hat{b}_{\omega} + e^{i\omega \eta}\hat{b}_{\omega}^\dagger \bigg)\,K_{i\omega}\big(m e^{\xi}\big)\sqrt{\sinh\pi \omega}\, {d \omega},
\end{align}
where the creation and annihilation operators obey the standard commutation relations:
\begin{align*}
    [\hat{b}^{}_{\omega},\hat{b}^\dagger_{\omega'}]=\delta(\omega-\omega'), \qquad  [\hat{b}^{}_{\omega},\hat{b}^{}_{\omega'}]=0.
\end{align*}
%
The so-called  Fulling vacuum \cite{ful1,ful2} is identified by the condition
\begin{equation}\label{fullingvac}
\hat b_\omega |0_R \rangle = 0,  \ \ \ \ \ \omega \geq 0.
\end{equation}
It is a pure state and the  corresponding two-point function  is given by
\begin{equation}
W_\infty(X_1,X_2)= \langle 0_R |\, \hat \varphi(X_1) \hat \varphi(X_2) |0_R \rangle =
\frac 1{\pi^2}\int_0^\infty e^{-i\omega (\eta_1-\eta_2) }  \,
 K_{i\omega}(m e^{\xi_1}) K_{i\omega}(m e^{\xi_2}) \,{\sinh \pi \omega}\ d\omega. \label{rindler1}
\end{equation}
The thermal equilibrium average of an operator ${\cal O}$ at  temperature  $T=\beta^{-1}$ is defined in quantum mechanics as follows:
\begin{equation}\label{defW}
    \langle\hat{\cal O}\rangle=\frac{\text{Tr}\,  e^{-\beta \hat{H}} \hat{\cal O}}{\text{Tr} \, e^{-\beta \hat{H}} },
\end{equation}
where $\hat{H}$ is the Hamiltonian of the system. This definition does not directly work in Quantum Field Theory but there are states having the same general properties expressed in terms of their analyticity  and periodicity properties in the complex time variable; they are known as the Kubo-Martin-Schwinger (KMS) states \cite{haag}.

In the Rindler wedge, a Wightman function having the KMS property may  be obtained by a generalized Bogoliubov tranformation \cite{ms,ms2} of the Fulling vacuum. In the case under consideration it is given by:
\begin{equation}
W_{\beta}(X_1(\eta_1,\xi_1),X_2(\eta_2,\xi_2)) =
\frac 1{ \pi^2}\int_0^\infty \left[ \frac{e^{-i\omega (\eta_1-\eta_2) }}{1-e^{-\beta\omega}} +    \frac{e^{i\omega (\eta_1-\eta_2)}}{ e^{\beta \omega}-1}\right]
K_{i\omega}(m e^{\xi_1}) K_{i\omega}(m e^{\xi_2}) {\sinh \pi \omega}\, d\omega. \label{propbetamassive}
\end{equation}
The two-point function  (\ref{propbetamassive}) is time-translation invariant (and therefore it provides an equilibrium state) and {\em respects the exchange symmetry} (\ref{symmetry}).
The  formal proof of the KMS periodicity {  makes use of the  exchange symmetry (\ref{symmetry})} but is otherwise straightforward.

%
%
When $\beta=
2\pi
$
an explicit  calculation of the integral shows that the LHS in Eq. \eqref{propbetamassive} can be extended to the whole complex Minkowski spacetime (minus the causal cut) and it is actually Poincar\'e invariant \cite{{Unruh:1976db},ms,ms2}:
\begin{align}
W_{2\pi}(X_1, X_2)=\frac{1}{2\pi}K_0\left(m \sqrt{-(X_1-X_2)^2}\right).
\end{align}
When $mL\to 0$ it has the standard ultraviolet (Hadamard behaviour) divergence  with the correct coefficient ${1}/{4\pi}$ :
\begin{align}
\label{k0log}
\frac{1}{2\pi}K_0\left(m\sqrt{-L}\right) \approx -\frac{1}{4\pi}\log(-m^2 \, L).
\end{align}
 Inside the Rindler wedge, the main contributions to the integral \eqref{propbetamassive} for  light--like separations come from high energies  $\omega \gg me^{\xi_{1,2}}$ ($\xi_{1,2}$ fixed) and the divergence does not depend on the temperature. 
This is true for any $\beta$.
However, when $\beta \neq 2\pi$ there are extra (anomalous) singularities at the  horizon ---
the boundary of the wedge, which of course is also light--like.
We will show this now.

When the  temperature is an integer multiple of {$ (2\pi)^{-1}$} a simple formula is available \cite{Akhmedov:2020qxd,Akhmedov:2019esv}:
\begin{align}
\label{2pin}
W_{\frac{2\pi}{N}} \left(X_1 , X_2\right)=\sum_{k=0}^{N-1} W_{2\pi} \left(X_1\left(\eta_1 - \frac{2\pi i\, k }{n},\,\xi_1\right) , \ X_2\left(\eta_2, \,\xi_2\right) \right).
\end{align}
Let us consider the simplest case  $\beta=\pi$:
\begin{align}
\label{proppi}
    W_{\pi}=\frac{1}{2\pi}K_0\left(m\sqrt{e^{2\xi_1}+e^{2\xi_2}-2e^{\xi_1+\xi_2}\cosh\Delta\eta}\right)+\frac{1}{2\pi}K_0\left(m\sqrt{e^{2\xi_1}+e^{2\xi_2}+2e^{\xi_1+\xi_2}\cosh\Delta\eta}\right).
\end{align}
Points of the   horizons may be  attained as follows:
\begin{eqnarray}
\label{coordinateshorizon}
\lim_{\lambda \to \pm\infty}
X(\lambda  ,\chi\mp \lambda) =\lim_{\lambda \to \infty}\left(
\begin{array}{c}
  e^{\chi \mp \lambda } \sinh  \lambda  \\
 e^{ \chi  \mp \lambda } \cosh  \lambda \\
\end{array}
\right)
\label{lorentz0}
= \frac 12 \left(
\begin{array}{c}
\pm  e^{ \chi } \\
 e^{  \chi }
\end{array}
\right)
\end{eqnarray}
The interval between two  points having the same coordinate $\lambda$ is spacelike; for instance
\begin{eqnarray}
\label{coordinateshorizon1}
L_{12}=\left(X_1(\lambda  ,\chi_1-\lambda) -X_2(\lambda  ,\chi_2-\lambda)\right)^2 = -e^{-2 \lambda} \left(e^{\chi_1}-e^{\chi_2}\right)^2<0;
\end{eqnarray}
furthermore
\begin{eqnarray}
 \left(X_1(\lambda -i \pi ,\chi_1-\lambda) -X_2(\lambda  ,\chi_2-\lambda)\right)^2 = -e^{-2 \lambda} \left(e^{\chi_1}+e^{\chi_2}\right)^2 = L_{12} + 4 e^{-2 \lambda} e^{\chi_1 + \chi_2}.
\end{eqnarray}
The first term in \eqref{proppi} is singular for any two light-like separated points  in the Rindler wedge.
When the two points are both approaching either  the future or  the past horizon also the second  term diverges, and it does exactly as the first term; when $\lambda \to +\infty$
\begin{align}
    W_{\pi}\big[X(\lambda  ,\chi_1-\lambda),X(\lambda  ,\chi_2-\lambda)\big] \approx -\frac{2}{4\pi}\log(-m^2 \, L_{12}), \ \ \text{as} \ \ \lambda \to +\infty.
\end{align}
Similarly for $\beta=\frac{2\pi}{N}$ and $ \lambda \to +\infty$.
\begin{align}\label{betaL}
    W_{\frac{2\pi}{N}}\big[X(\lambda  ,\chi_1-\lambda),X(\lambda  ,\chi_2-\lambda)\big] \approx -\frac{N}{4\pi}\log(- m^2 \, L_{12}) = -\frac{1}{2\beta}\log(-m^2 \, L_{12}).
\end{align}
In the horizon limit \eqref{coordinateshorizon} the dominant contribution to the integral (\ref{propbetamassive}) comes from the infrared region  $\omega \to 0$. Using appendix A and asymptotic form of the modes near horizon one can show that \eqref{betaL} remains true for general $\beta$. The calculation is similar to the one preformed in \cite{Akhmedov:2020qxd}.
Such a dependence of the coefficient of the singularity at light--like separation (at the horizon) implies that the thermal state  cannot be continued to the entire Minkowski space--time.

It is possible to introduce more general time translation invariant (at tree--level) states by letting the temperature depend on the energy:

\begin{equation}
{\cal W} (X_1,X_2) =
\frac 1{ \pi^2}\int_0^\infty \left[ \frac{e^{-i\omega (\eta_1-\eta_2) }}{1-e^{-\beta(\omega)\, \omega}} +    \frac{e^{i\omega (\eta_1-\eta_2)}}{ e^{\beta(\omega) \, \omega}-1}\right]
K_{i\omega}(m e^{\xi_1}) K_{i\omega}(m e^{\xi_2}) {\sinh \pi \omega}\, d\omega. \label{propbetamassive2}
\end{equation}
These states also respect the exchange symmetry (\ref{symmetry}).
However, when $m\neq 0$ it is not possible to disentangle two independent temperatures for the left-- and right--movers. That is because the modes $e^{-i\omega t} \, K_{i\omega}$ are normalizable linear combinations of left-- and right--movers. We will see below that for de Sitter and Schwarzschild fields the situation is different and such a possibility does exist.



\subsection{General dimension}
If there are $d$ extra flat transverse spatial dimensions $\vec{x}$ then the Rindler metric is:

\begin{align}
    ds_d^2=e^{2\xi}\big(d \eta^2-d\xi^2\big)-d\vec{x}^2.
\end{align}
The modes can be represented as $\varphi(\eta,\xi,\vec{x})= e^{i \vec{k}\vec{x}} \varphi_{\vec{k}} (\eta,\xi) $ where $\varphi_{\vec{k}} (\eta,\xi) $ obeys Eq. (\ref{eqmassive}) with the effective mass $m^2 + k^2$. Therefore the field operator can be expanded as

\begin{align}
\label{operatormassiveRindlerD}
    \hat{\varphi}(\eta,\xi,\vec{x}) =\int_{-\infty}^{+\infty} \frac{d^dk}{(2\pi)^\frac{d}{2}} \int_{0}^{+\infty} \frac{d \omega}{\pi}\sqrt{\sinh\pi \omega}\bigg[e^{-i\omega \eta+i\vec{k}\vec{x}}\hat{b}_{\omega,\vec{k}}^{}  +e^{i\omega \eta- i\vec{k}\vec{x}}\hat{b}_{\omega,\vec{k}}^\dagger\bigg]K_{i\omega}\big(\sqrt{m^2+k^2} e^{\xi}\big).
\end{align}
The Wightman function at temperature $\beta$ is as follows:
\begin{eqnarray}\label{genericbetarind}
&& W_{\beta}(X_1,X_2)=\cr&& =
     \int_{-\infty}^{+\infty} \frac{d^dk}{(2\pi)^d} \int_{-\infty}^{+\infty} \frac{d\omega}{\pi^2}  \frac{\sinh(\pi \omega)}{1-e^{-\beta \omega}}   e^{-i \omega (\eta_1-\eta_2)}e^{i\vec{k} (\vec{x}_1-\vec{x}_2)} K_{i\omega}\big(\sqrt{m^2+k^2} e^{\xi_1}\big)K_{i\omega}\big(\sqrt{m^2+k^2} e^{\xi_2}\big) \cr &&
\end{eqnarray}
Enforcing Poincar\'e invariance gives $\beta = 2\pi$ \cite{ms}; this is the well-known Bisognano--Wichmann theorem, valid also for interacting quantum fields \cite{bisognano}.
The anomalous divergence on the horizons for generic $\beta \neq 2\pi$ goes precisely as in the  previous section.




\subsection{Stress energy tensor in 2D}

Here we complete the discussion of the  massive scalar field in 2D Rindler spacetime by examining the renormalized stress-energy tensor at various temperatures (some technical details can be found in  appendix \ref{appendixD}). 
To set up the notations let us summarise
the standard expression resulting from point splitting regularization  in the Poincar\'e invariant case $\beta = 2\pi$:
\begin{eqnarray}
 &&  \langle     T_{VV}\rangle_{2\pi}  = -\frac{t_V t_V}{4\pi\epsilon^2 }, \quad
   \langle   T_{UU}\rangle_{2\pi}  =  -\frac{t_U t_U}{4\pi\epsilon^2 },\cr
 &&   \langle   T_{VU}\rangle_{2\pi} =\langle   T_{UV}\rangle_{2\pi} =-\frac{e^{V-U}}{8 \pi } m^2 \left[ \gamma_e  + \log ( m ) +\log \left(\epsilon\sqrt{t_\alpha t^\alpha}\right) \right],
\end{eqnarray}
where $t_\mu$ is the vector separating the two points of the Wightman function \eqref{propbetamassive}. 

The above expressions lead
to the covariantly conserved stress--energy tensor \cite{birreldavies}:
%
\begin{align}
\label{2piset1}
\langle   :T_{\mu\nu}: \rangle_{2\pi}=-\frac{1}{4 \pi } m^2 \left[ \gamma  + \log ( m )  \right] g_{\mu\nu},
\end{align}
where $\gamma$ is the Euler-Mascheroni constant. This is obviously related to the expectation value in Minkowski space by the coordinate transformations \eqref{coordinatesuv}.

Similarly, for  $\beta = 2\pi/N$   point splitting  regularization in \eqref{2pin} gives
%
\begin{align}
\langle   :T_{\mu\nu: }\rangle _{\frac{2\pi}{N}}=
\sum_{n=1}^{N-1}  \frac{m^2}{4} e^{V - U}
     K_2\left( 2 m e^{\frac{V - U}{2}} \sin \left( \frac{n \pi}{N}\right) \right)
    \begin{bmatrix}
1    &   0     \\
0 &  1
\end{bmatrix} +
\cr  +
\left(  -\frac{1}{4 \pi } m^2 \left[ \gamma  + \log ( m )  \right]+\frac{m^2}{2} \sum_{n=1}^{N-1} K_0\left( 2 m e^{\frac{V - U}{2}} \sin \left( \frac{n \pi }{N} \right)\right)\right)  g_{\mu\nu},\label{betaNsttenz}
\end{align}
where $K_0(x)$ and $K_2(x)$ are MacDonald functions. Violation of Poincar\'e invariance is manifest.

Near the horizon this expression simplifies to:
\begin{eqnarray} \label{TbetaN}
\langle  :T_{\mu\nu }:\rangle _{\frac{2\pi}{N}}=
\frac{1}{24}\left(N^2-1\right)
    \begin{bmatrix}
1    &   0     \\
0 &  1
\end{bmatrix} + {\mathcal O}(e^{V-U}),
\end{eqnarray}
while at the spatial infinity it gives:
\begin{align*}
\langle  :T_{\mu\nu }:\rangle _{\frac{2\pi}{N}} \approx -\frac{1}{4 \pi } m^2 \left[ \gamma_e  + \log ( m )  \right] g_{\mu\nu},
\end{align*}
which coincides with the $\beta = 2 \pi$ case.
These two types of asymptotic behaviour of the stress energy tensor are regular. Furthermore, the second one does not depend on $\beta$.
On the other hand, the expectation value of the mixed components of stress--energy tensor  $T_\mu^\nu$ diverge at the horizon. 
%
For generic values of $\beta$, when both points 
in \eqref{genericbetarind} are taken to the horizon we get (see appendix B)
\begin{multline}
 W_{\beta}(X^+,X^-) \approx
 \int_{-\infty}^{\infty}\frac{d\omega}{\pi\omega}\frac{e^{-\frac{i\omega}{2}
 {(V^+ +U^+-V^--U^-)}}}{1-e^{-\beta\omega}}\sin\left(\omega\log(me^{\frac{(V^+-U^+)}{2}}/2) +\arg\Gamma (1-i\omega)\right) \times \\
 \times \sin\left(\omega\log(me^{\frac{(V^--U^-)}{2}}/2)+\arg\Gamma (1-i\omega)\right).
 \end{multline}
 The expectation value may be obtained by taking into account
\begin{multline}
 \partial _{V_+}\partial _{V_-}  W_{\beta}(X^+,X^-)=   \int_{-\infty}^{\infty}\frac{d\omega}{4\pi} \frac{\omega}{1-e^{-\beta\omega}} e^{-i\omega (V^+ -V^-)}= \\
 =-\frac{1}{4 \pi (V^+-V^-)^2 }+\frac{\pi}{12 \beta^2}
 =-\frac{t_V t_V}{4\pi\epsilon^2 }+\frac{1}{24\pi}+\frac{\pi}{12 \beta^2}.
\end{multline}
At the horizon for arbitrary temperatures we get
\begin{align}
\langle   :T_{\mu\nu }:\rangle _\beta=
\frac{1}{24}\left( \left(\frac{2 \pi}{\beta}\right)^2-1\right)
    \begin{bmatrix}
1    &   0     \\
0 &  1
\end{bmatrix} + {\mathcal O}(e^{V-U})
    \end{align}
to be compared  with \eqref{TbetaN}.

\subsection{Massless case}

We complete this section with a few remarks on the massless case, necessary to understand the novel features of the de Sitter case which are presented  in the next section. The discussion is kept short. Details  will be fully discussed elsewhere.

The massless  Klein-Gordon equation in the two--dimensional Rindler spacetime is

\begin{equation}
    \Box \phi = \partial_\eta^2\phi  -\partial_\xi^2 \phi = 0.
\end{equation}
As regards the vacuum two-point function as in the Minkowskian case  \cite{Wightman,klaiber} one would set in Fourier space
 \begin{equation}
\widetilde W_0(k)  = 
 \theta(k^0)\delta(k^2)  = \widetilde W_R(k)  + \widetilde W_L(k) = \frac{1}{k^+}  \theta(k^+) \delta(k^-) + \frac{1}{k^-} \theta(k^-) \delta(k^+)
 \end{equation}
where we introduced the lightcone variables  $k^\pm = k^0\pm k^1$.
Unfortunately  $\theta(k^0)$ is not a multiplier for $\delta(k^2)$. The standard regularization (see e.g. \cite{Wightman,klaiber,gelfand,strocchi}) involves an arbitrary infrared regulator having the dimension of a mass:
\begin{eqnarray}
W_R(\eta,\xi )= W_R(U)
=- \frac 1 {4\pi} { \log \left({i \mu  (U -i \epsilon)}\right)}, \\
W_L(\eta,\xi) =W_L(V)
= - \frac 1 {4\pi} { \log \left({i \mu (V-i \epsilon)}\right)}.
\end{eqnarray}
We see here that $ W_R$ (resp. $ W_L$) is analytic in the lower half-plane of the complex variable $U$
 (resp.  $V$).
We may therefore introduce the regularized thermal massless right two-point function as the following formal series
  \begin{eqnarray}
W_{R,\beta} (x^-) 
=
\sum_{n=0}^{\infty} W_R(U -i n \beta ) + \sum_{n=1}^{\infty} W'_R(U + in \beta)
 \end{eqnarray}
and similarly for
$
W_{L,\beta} (V)$. 
The total two-point function is the sum
    \begin{eqnarray}
W_{0,\beta} (\eta,\xi)  = W_{L,\beta} (V)+W_{R,\beta} (U).
 \end{eqnarray}
But the left and right movers are independent fields and  may have independent temperatures. We may thus formally introduce the states characterized by functions of the form

\begin{eqnarray}
W_{0,\beta_L \beta_R} (x)  = W_{L,\beta_L} (x^+)+ W_{R,\beta_R} (x^-).
\end{eqnarray}
This choice of course  does not change the  commutators that do not depend on the temperatures.

In conclusion we may introduce two independent temperatures for the left and right moving fields

\begin{equation}
    \phi = \phi_R+\phi_L
\end{equation}
by leaving of course the commutator untouched.
But this possibility in Rindler space exist only for the massless field and not for the massive one.

\section{Static patch of de Sitter space--time}
\label{static}

\subsection{Geometry, modes and Wightman functions}

\begin{figure}[!h]
\centering
\includestandalone[width=0.8\textwidth]{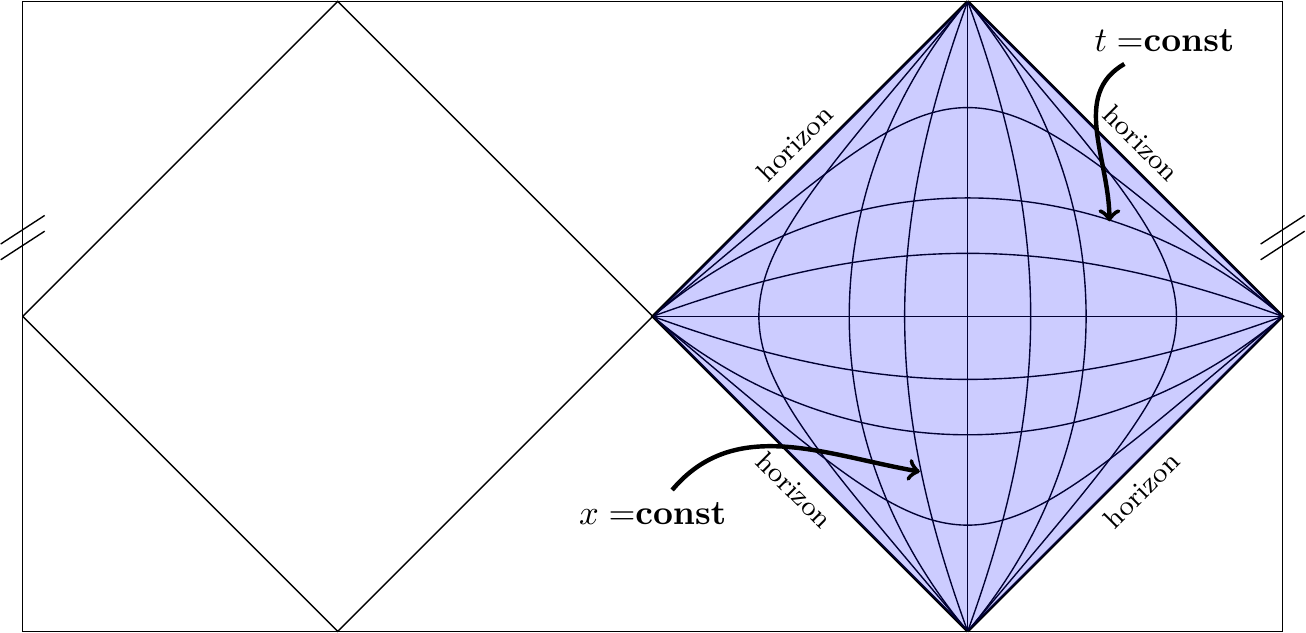}
\caption{Penrose diagram of the de Sitter manifold with Cauchy surfaces of different patches. The Static patch is bordered by a bifurcate Killing horizon.} \label{fig2}
\label{staticpic}
\end{figure}

The two-dimensional de Sitter space can be most easily visualized as the one-sheeted  hyperboloid embedded in a three dimensional ambient Minkowski space:
\begin{align}
dS_2 = \{X\in {\bf R}^3, \ \  X^\mu X_\mu=X_{0}^2-X_{1}^2-X_{2}^2=-R^2\} . 
\end{align}
$X^{\mu}$ denote the coordinates of a given Lorentzian frame  of the ambient spacetime;  we set the radius $R$ of  the de Sitter space equal to one. A suitable  coordinate system for the static patch is
\begin{equation}
\label{coordinates}
 X\left(t , x \right)=\begin{cases}
   X^0=  \sinh  t  \ { \rm sech}\  x \\
   X^1= \tanh x  = u \\
   X^2=  \cosh  t \ {\rm sech}\  x
 \end{cases},
 \qquad t \in(-\infty,\infty), \ x \in(-\infty,\infty).
\end{equation}
%
The static coordinates   cover a causal diamond  of the entire two dimensional de Sitter space (see Fig. \ref{fig2}; see also  \cite{Akhmedov:2020qxd} for a full description); we refer to it as the static patch or the Rindler--de Sitter wedge.  The metric and the massive scalar Klein-Gordon equation  in these coordinates are written as follows:
\begin{eqnarray}
&& ds^2 = \frac{dt^2-dx^2}{\cosh x {}^2}, \\
\label{sdseq}
&& \partial_{t}^2 \varphi - \partial_x^2  \varphi+ \frac{m^2\varphi }{\cosh^2 x} =0.\nonumber
\end{eqnarray}
 The scattering eigenfunctions of the Schr\"odinger  operator with potential $m^2/\cosh^2 x$ \cite{Akhmedov:2020qxd}
\begin{align*}
\psi_\omega(x)=\sqrt{\sinh(\pi\omega)} \,  \Gamma\left(\frac{1}{2}+i\mu-i\omega\right) \, \Gamma\left(\frac{1}{2}-i\mu-i\omega \right) \, \mathsf{P}^{i\omega}_{-\frac{1}{2}+i\mu}(\tanh x),    \ \ \  \mu^2=m^2-\frac{1}{4}
\end{align*}
provide  two modes   for each energy level $\omega$, namely $e^{-i \omega t} \, \psi_\omega(\pm x)$.
The asymptotic behaviour of the modes at  $x\to \infty$ is governed by
\begin{eqnarray}
\label{plus}
 && \mathsf{P}^{i\omega}_\ind\left(\tanh x\right) \underset{x\to\infty}{\approx} \frac{e^{i\omega x}}{\Gamma(1-i\omega)},
\\
\label{minus}
&& \mathsf{P}^{i\omega}_\ind\left(-\tanh x\right) \underset{x\to\infty}{\approx}  \bigg[ \frac{\Gamma\big(-i\omega\big) e^{-i\omega x}}{\Gamma\big(\frac12+i\mu-i\omega\big)\Gamma\big(\frac12-i\mu-i\omega\big)}+\frac{\cosh(\mu\pi)\Gamma\big(i\omega\big) \, e^{i\omega x}}{\pi }\bigg];
\end{eqnarray}
this shows that  $e^{-i \omega t} \, \psi_\omega(x)$ is asymptotically right--moving  and $e^{-i \omega t} \, \psi_\omega(-x)$ left--moving.

The  expansion of the field operator written in terms of the above modes {\em naturally splits into two commuting fields: a left mover ${\phi_L}$ and a right mover ${\phi_R}$ for all values of the mass $m$}:
\begin{eqnarray}
\label{fieldoperator}
 {\phi}(t,x)&=&{\phi_R}(t,x)+ {\phi_L}(t,x), \cr && \cr
 {\phi_R}(t,x)&=&   \frac{1}{2\pi}\int_{0}^\infty   \left[e^{-i \omega t} \psi_\omega(x)a_\omega  +e^{i \omega t} \psi^*_\omega(x) a^\dagger_\omega \right] {d\omega},\cr  && \cr {\phi_L}(t,x)&=&  \frac{1}{2\pi}\int_{0}^\infty   \left[ e^{-i \omega t} \psi_\omega(-x) b_\omega +e^{i \omega t} \psi^*_\omega(-x) b^\dagger_\omega \right] d\omega;
\end{eqnarray}
the  ladder operators  obey the standard commutation relations
\begin{align*}
\big[a_{\omega_1}, a^\dagger_{\omega_2} \big]=\delta(\omega_1-\omega_2), \ \ \ \big[b_{\omega_1}, b^\dagger_{\omega_2} \big]= \delta(\omega_1-\omega_2), \qquad \big[a_{\omega_1}, b_{\omega_2} \big]=\big[a_{\omega_1}, b^\dagger_{\omega_2} \big]=0.
\end{align*}
It maybe worthwhile to stress that the above separation into left and right movers is only possible in the static coordinate system because of the symmetry of the effective potential and, once more, it holds true for massive fields. { Here the left and right moving modes only asymptotically depend on one of the two lightcone variables $t\pm x$ near the corresponding side of the horizon.}

In \cite{Akhmedov:2020qxd} we  constructed  general time translation invariant states
\begin{align}
\label{bithermalst}
    \langle\hat{a}_\omega^\dagger \hat{a}_{\omega'}^{} \rangle=\delta(\omega-\omega')\frac{1}{e^{\beta_R(\omega) \,\omega}-1} \qquad  \text{and} \qquad \langle\hat{b}_\omega^\dagger \hat{b}_{\omega'}^{} \rangle=\delta(\omega-\omega')\frac{1}{e^{\beta_L(\omega)\, \omega}-1}.
\end{align}
 We gave in particular a full treatment for states of arbitrary global (inverse)  temperature
\begin{equation}
    \beta_L(\omega)= \beta_R(\omega)=\beta
\end{equation}and provided new integral representations for their correlation functions. Taking inspiration from the consideration of Unruh state for black holes,  we enlarge that study  and consider  different global  temperatures for  the left and the right--moving modes:
\begin{equation}
    \beta_L(\omega)=\beta_L, \ \ \  \beta_R(\omega)=\beta_R.
\end{equation}
The Wightman function is the sum of two contributions
 \begin{equation}
 \label{propstatic}
   W_{\beta_L\beta_R}(X_1,X_2 )
   = W_{L,\beta_L}(X_1,X_2 )+W_{R,\beta_R}(X_1,X_2 ),
 \end{equation}
 where
  \begin{eqnarray}
 \label{propstatic2a}
   W_{L,\beta}(X_1,X_2 )
   =\int_{0}^{\infty}\frac{d\omega  }{4\pi^2}   \left[   e^{-i\omega (t_1-t_2)}   \frac{\psi_{\omega}(-x_1) \psi_{\omega}^*(-x_2)}{1-e^{-\beta \omega }}  +
e^{i\omega( t_1- t_2)}     \frac{\psi^*_{\omega}(-x_1) \psi_{\omega}(-x_2)}{e^{\beta \omega }-1} \right],
\\   W_{R,\beta}(X_1,X_2 )
   =
   \int_{0}^{\infty}\frac{d\omega  }{4\pi^2}   \left[   e^{-i\omega (t_1-t_2)}   \frac{\psi_{\omega}(x_1) \psi_{\omega}^*(x_2)}{1-e^{-\beta \omega }}  +
e^{i\omega( t_1- t_2)}     \frac{\psi^*_{\omega}(x_1) \psi_{\omega}(x_2)}{e^{\beta \omega }-1} \right].
 \label{propstatic2} \end{eqnarray}
The formal proof of the KMS periodicity property goes as follows:
\begin{eqnarray}
&& W_{R,\beta}(X_2(t_2,x_2),X_1(t_1,x_1)) =\frac 1{4 \pi^2}\sum_{n=0}^\infty\int_0^\infty  {e^{-i\omega (t_2-t_1-i n \beta) }}\psi_{\omega}(x_2) \psi_{\omega}^*(x_1)
d\omega +\cr && \cr
&& +\frac 1{4 \pi^2}\sum_{n=1}^\infty\int_0^\infty    {e^{i\omega (t_2-t_1+in\beta)}}\psi^*_{\omega}(x_2) \psi_{\omega}(x_1) \, d\omega= W_{R,\beta}(X_1(t_1-i\beta,x_1),X_2(t_2,x_2))\cr && \end{eqnarray}
 There holds the exchange symmetry
  \begin{equation}
 \label{symmetry22}
   W_{R,\beta_R}(X_1(t_1,x_1),X_2(t_2,x_2) ) =
   W_{L,\beta_R}(X_1(t_1,-x_1),X_2(t_2,-x_2) )
 \end{equation}
%
When $\beta_L = \beta_R = 2 \pi$ the Wightman function  (\ref{propstatic}) respects the de Sitter isometry \cite{HG,Akhmedov:2020qxd,fhkn,sewell,bgm,bm,nt}, i.e. it is a function of the complex de Sitter invariant variable

$$
\zeta = - \frac{\cosh(t_2-t_1) + \sinh x_1 \sinh x_2}{\cosh x_1 \cosh x_2}
$$
with the locality cut on the negative reals. The variable $\zeta$ and
the geodesic distance $L$ are related as follows:
$\zeta = - \cosh L \ $  for time-like geodesics,   $\zeta = \cos L\ $ for space-like ones;  $\zeta = - 1$ for light-like separations.

Let us consider now the  behavior at the horizon of Eq. (\ref{propstatic}). Points of the right (left) future horizon are obtained in the following limit 
\begin{eqnarray}
\label{coordinateshorizonds}
\lim_{\lambda \to + \infty}
X(\lambda  ,\pm (\lambda - \chi)) =\lim_{\lambda \to +\infty}\left(
\begin{array}{l}
{ \rm sech\, }(\lambda -\chi) \sinh  \lambda
 \\
 \pm  \tanh  (\lambda-\chi )
 \\  { \rm sech\, }(\lambda-\chi ) \cosh  \lambda  \\
\end{array}
\right)
= \left(
\begin{array}{r}
 e^{ \chi } \\ \pm1 \\
 e^{  \chi }
\end{array}
\right)
\end{eqnarray}
Points of the left (right)  past horizon are obtained in the limit $\lambda  \to -\infty$ of the above expression. In all cases
the interval between two points having the same finite coordinate $\lambda$ is spacelike:
\begin{eqnarray}
\label{coordinateshorizon1}
L_{12}= -\frac{2 (\cosh (\chi_1 -\chi_2 )-1)}  { \cosh (\lambda -\chi_1 ) \cosh (\lambda -\chi_2 ) }       <0,
\end{eqnarray}
becoming light--like only in the limit $\lambda \to \pm \infty$.

Using the asymptotics of the modes \ref{plus} and \ref{minus}, and the eq. \eqref{helpfull}, one can obtain the behaviour of $W_{R,\beta_R}$ and $W_{L,\beta_L}$ separately at e.g. the right side of the horizon. As we can see from Eq. \eqref{plus}, in this region $W_{R,\beta_R}$ depends only on the difference $x_1-x_2$, which does  not grow when both points are taken to the same side of the horizon. It means, that this contribution to the Wightman function is regular near the right side of the horizon. At the same time, in the same region $W_{L,\beta_L}$ depends on the both $x_1-x_2$ and $x_1+x_2$, as we can see from Eq. \eqref{minus}. The latter sum is infinitely growing near the horizon. As the result, using \eqref{helpfull} one obtains that:

\begin{align}
     W_{\beta_L\beta_R}\Big(X(\lambda  ,\chi_1-\lambda) ,  X(\lambda  ,\chi_2-\lambda) \Big)       \approx W_{L,\beta_L}\Big(X(\lambda  ,\chi_1-\lambda) ,  X(\lambda  ,\chi_2-\lambda) \Big) \approx \frac{1}{\beta_L}\lambda, \qquad \lambda \to +\infty.
\end{align}
Behavior near the left horizon can be found from the symmetry from Eq. \eqref{symmetry22}, which implies that
$$
W_{\beta_L\beta_R}\left(X(t_1,x_1 ), X( t_2, x_2)\right)= W_{\beta_R\beta_L}\left(X(t_1,-x_1 ), X( t_2, -x_2)\right).
$$
Hence parity $x\to -x$ plus  rearrangement of temperatures $\beta_L\leftrightarrow\beta_R$ leave the two-point-function invariant.
As a result, it follows that for  $\lambda \to -\infty$
\begin{align}
     W_{\beta_L\beta_R}\Big(X(\lambda  ,\chi_1-\lambda) ,  X(\lambda  ,\chi_2-\lambda) \Big)      \approx W_{R,\beta_R}\Big(X(\lambda  ,\chi_1-\lambda) ,  X(\lambda  ,\chi_2-\lambda) \Big) \approx\frac{1}{\beta_R}|\lambda|.
\end{align}
The light--like singularity at the horizons depends on the state of the theory. In particular, at the right horizon it depends only on $\beta_R$ while at the left horizon it depends on $\beta_L$. This shows that such a peculiar behavior is present due to the interplay between the waves that are falling down and reflected from the $m^2/\cosh^2 x$ potential.

\subsection{General dimension}

The $(D+1)$-embedding coordinates and the invariant scalar product for the  $D$--dimensional static patch are given by
\begin{align}
    X_0 = \sinh(t) \, \text{sech}(x),  \quad X_i = \tanh(x) \, \vec{y}_i, \quad X_D = \cosh(t) \, \text{sech}(x), \quad \vec{y}_i\vec{y}_i=1,
\\
    Z = \eta_{\mu\nu} \, X_1^\mu \, X_2^\nu = -\frac{\cosh(t_2-t_1) + \vec{y}_1\cdot  \vec{y}_2 \  \sinh x_1 \, \sinh x_2 \, }{\cosh x_1 \, \cosh x_2}.
\end{align}
The Bunch-Davies Wightman function \cite{HG,thir,nach,tagirov,ss,Bunch,bgm,bm}  corresponding to the inverse temperature $\beta_L = \beta_R = 2\pi$  \cite{HG,Akhmedov:2019esv,sewell,bgm,bm} is given by
\begin{align}
W_{2\pi}(Z)=\frac{\Gamma\left(\frac{D-1}{2} + i \, \mu\right)\Gamma\left(\frac{D-1}{2} - i \, \mu\right)}{ 2 (2\pi)^{\frac{D}{2}}}  (Z^2-1)^{-\frac{D-2}{4}} P^{-\frac{D-2}{2}}_{-\frac{1}{2}+ i\mu}(Z),
\end{align}
where $\mu = \sqrt{m^2 - (D-1)^2/4}$.
It has the standard Hadamard  singularity near $Z=-1$.

Points of the future and past horizons are attained in the following limits 
\begin{eqnarray}
\label{coordinateshorizonds}
\lim_{\lambda \to \pm \infty}
X(\lambda  , (\lambda - \chi)) =\lim_{\lambda \to \infty}\left(
\begin{array}{l}
{ \rm sech\, }(\lambda -\chi) \sinh  \lambda
 \\
   \tanh  (\lambda-\chi ) \vec y
 \\  { \rm sech\, }(\lambda-\chi ) \cosh  \lambda  \\
\end{array}
\right)
\label{lorentz0}
= \left(
\begin{array}{l}
\pm e^{ \chi } \\ \pm \vec y \\
 e^{  \chi }
\end{array}
\right)
\end{eqnarray}
Two events on the horizons are spacelike separated unless   $\vec{y}_1=\vec{y}_2$. As in Eq. \eqref{2pin} for $\beta=\frac{2\pi}{N}$ in the horizon limit one gets:

\begin{align}
      W_{\frac{2\pi}{N}}(\lambda\to + \infty) \approx  N \, W_{2\pi}(\lambda \to + \infty) \approx - N \,  \frac{\Gamma\Big(\frac{D-2}{2}\Big)}{2^{2+(D-2)\frac{3}{2}}\pi^\frac{D}{2}}e^{(D-2)\lambda}.
\end{align}
As in Rindler space   the singularity of the propagator on the horizon depends on the  temperature.

\subsection{Stress-energy tensor in 2D}

Let us introduce the light-cone coordinates of the static patch:
\begin{eqnarray}
&&    V=t+x,\qquad U=t-x, \nonumber
\\ \label{metric}
&& ds^2=\frac{1}{\cosh^2 (\frac{V-U}{2})  }dUdV \equiv C(U,V) dU dV.
\end{eqnarray}
To set up  notations let us discuss first  the de Sitter invariant case $\beta = 2\pi$.
When the two arguments of the Wightman function are taken very close to each other, one has that
\begin{align}
W _{2\pi} (X_+, X_-) \approx - \frac{1}{4\pi}\left( H_{-\frac{1}{2}+i\mu}+ H_{-\frac{1}{2}-i\mu}+\log\left[\frac{(V_--V_+)(U_+-U_-)}{4 \cosh^2(V-U)}\right]\right),
\end{align}
where $H_{-\frac{1}{2}+i \mu}=\psi\left(\frac{1}{2}+i \mu\right)+ \gamma_e
$ are the harmonic numbers;
the definitions of $X_\pm$, $V_\pm$ and $U_\pm$ can be found in appendix B. Since
\begin{align}
 \partial _{V_+}\partial _{V_-}W_{2\pi}(Z)
  \approx -\frac{1}{4\pi(V_+-V_-)^2}+\frac{1}{48 \pi},
\quad {\rm and} \quad
 \partial _{U_+}\partial _{U_-}W_{2\pi}(Z)
  \approx -\frac{1}{4\pi(U_+-U_-)^2}+\frac{1}{48 \pi}.
 \end{align}
the covariant point splitting regularization gives
\begin{eqnarray}
 \langle   T_{UV}\rangle_{2\pi} =-\frac{m^2}{8 \pi \cosh^2 (\frac{V-U}{2} )}\left( \psi\left(\frac{1}{2}+i \mu\right)+\psi\left(\frac{1}{2}-i \mu\right)+ 2 \gamma_e+\log\left[\epsilon^2 \ t_\alpha t^\alpha\right]\right),\\  \langle    T_{UU}\rangle_{2\pi}  = -\left(\frac{1}{4\pi\epsilon^2 (t_\alpha t^\alpha)}+\frac{R}{24 \pi}\right) \frac{t_U t_U}{t_\alpha t^\alpha}, \\
   \langle   T_{VV}\rangle_{2\pi}  =-\left(\frac{1}{4\pi\epsilon^2 (t_\alpha t^\alpha)}+\frac{R}{24 \pi}\right) \frac{t_V t_V}{t_\alpha t^\alpha}.
\end{eqnarray}
After regularization we obtain the well known answer  \cite{Bunch}
\begin{align}
\label{2piset}
\langle   :T_{\mu\nu}:\rangle_{2\pi}=-\frac{1}{4 \pi } m^2 \left[ \psi\left(\frac{1}{2}+i \mu\right)+\psi\left(\frac{1}{2}-i \mu\right)+ 2 \gamma_e  \right] g_{\mu\nu} +  \frac{R}{48 \pi } g_{\mu \nu}
\end{align}
The expectation value of the stress--energy tensor with two  temperatures $\beta_L$ and $\beta_R$ can be obtained starting from Eq. \eqref{propstatic}. The most interesting case in this situation is the near horizon limit.
For instance, close to right horizon 
a lengthy but not difficult calculation gives
\begin{align}
    \partial_{V^+} \partial_{V^-} W_{\beta_L \beta_R} (X^+, X^-) \approx \int^{\infty}_{-\infty} \frac{d \w}{ 4 \pi } \frac{\w}{e^{\beta_L \w}-1} e^{i \w (V^+-V^-)} = \frac{\pi}{12 \beta_L^2} - \frac{1}{4\pi} \frac{1}{(V^+-V^-)^2}
\end{align}
and
\begin{multline} \label{desitter_integral}
    \partial_{U^+} \partial_{U^-} W_{\beta_L \beta_R} (X^+, X^-) \approx \int^{\infty}_{-\infty} \frac{d \w}{ 4 \pi }  \frac{\w \sinh^2 \pi \w \ e^{i \w (U^+-U^-)}}{\cosh \pi (\w- \mu) \cosh \pi (\w+\mu)}  \bigg[ \frac{1}{e^{\beta_R \w}-1} + \frac{\cosh^2 \mu \pi}{\sinh^2 \pi \w} \frac{1}{e^{\beta_L}-1} \bigg]
\end{multline}
%
The above expressions simplify when the temperatures of the left-- and right--movers coincide: $\beta_R = \beta_L = \beta$. Then the regularized stress-energy tensor in the near horizon limit takes the form:
\begin{align} \label{set_static}
    \langle :T_{\mu \nu}: \rangle \approx \Theta_{\mu \nu} + \frac{R}{48 \pi } g_{\mu \nu},
\end{align}
where 
\begin{align*}
    \Theta_{UU} &= -\frac{1}{12 \pi} C^{1/2} \partial^2_U C^{-1/2}+ \frac{\pi}{12 \beta^2} = \frac{\pi}{12} \bigg( \frac{1}{\beta^2} - \frac{1}{(2\pi)^2} \bigg) ,\\
    \Theta_{VV} &= -\frac{1}{12 \pi} C^{1/2} \partial^2_V C^{-1/2} + \frac{\pi}{12 \beta^2}=\frac{\pi}{12} \bigg( \frac{1}{\beta^2} - \frac{1}{(2\pi)^2} \bigg) ,\\
    \Theta_{UV} &= \Theta_{VU} =0.
\end{align*}
 de Sitter covariance is recovered only when $\beta = 2\pi$.


\section{Schwarzschild black hole}
\label{SchwarzschildVARIANT2}
\begin{figure}[!h]
\centering
\includestandalone[width=0.8\textwidth]{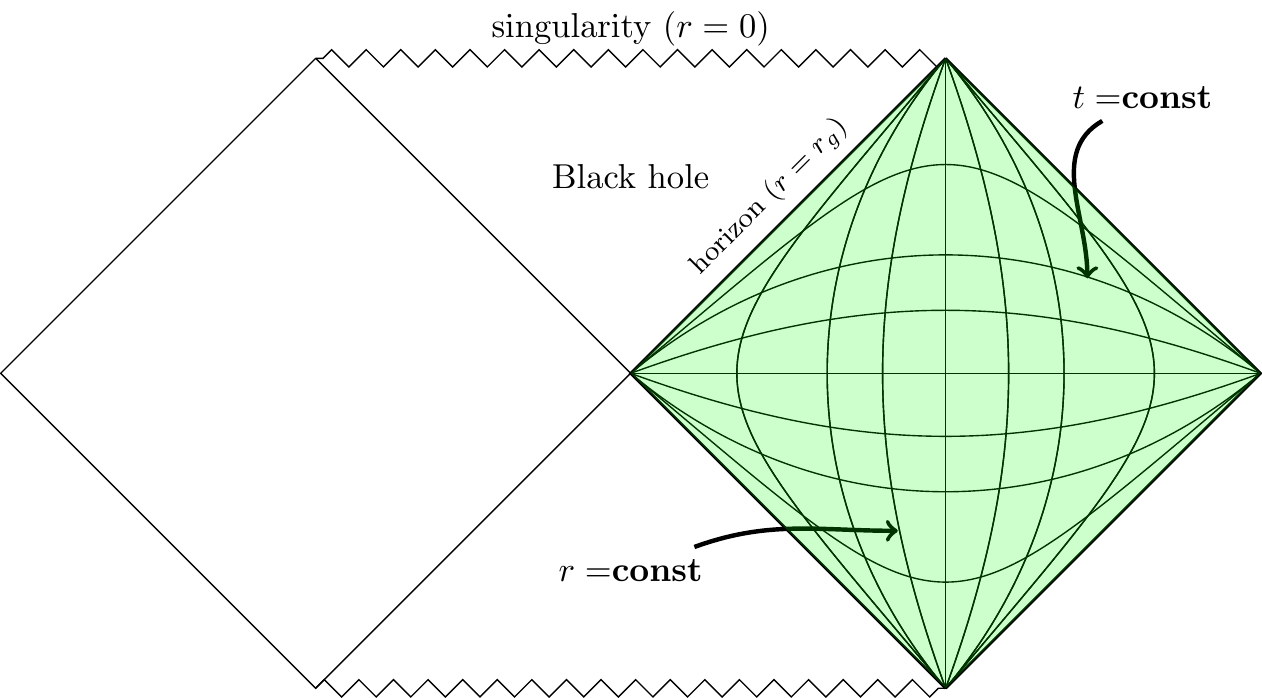}
\caption{Penrose diagram of the Schwarzschild black hole.}
\label{Schpic}
\end{figure}
Here we consider the radial part of the Schwarzschild metric (we call it the two--dimensional black hole):
\begin{equation}
ds^2 = \left(1 - \frac{r_g}{r}\right) \, dt^2 - \frac{dr^2}{1 - \frac{r_g}{r}} =  ds^2 = \bigg[1-\frac{r_g}{r(r^*)}\bigg]\Big(dt^2-d{r^*}^2\Big).\label{BHmetr}
\end{equation}
The tortoise coordinate 
\begin{align}
    r^*=r+r_g\log\Big(\frac{r}{r_g}-1\Big), 
\end{align}
is such that $r^*\approx r$ when $r\to+\infty$ while   $r^* \rightarrow -\infty$ when $r\to r_g$;
near the horizon the metric looks like the Rindler's one  \eqref{metricRindler} :
\begin{align}
\label{Schmetrichor}
   ds_{\text{nh}}^2\approx e^\frac{r^*}{r_g}\Big(dt^2-d{r^*}^2\Big)
\end{align}
with the acceleration $\alpha=\frac{1}{2r_g}$.
%
%


\subsection{Modes and Wightman function} \label{modwight}
In the tortoise coordinates the massive  Klein-Gordon field equation,

\begin{equation}
\label{Scheq}
\partial^2_t \varphi- \partial^2_{r^*} \varphi +m^2 g_{00}  \, \varphi=0,
\end{equation}
is such that the mass term in \eqref{Scheq} vanishes at the horizon where $g_{00}$ does vanish.
By separating the variables  $\varphi(t,r^*)=e^{i\omega t}\varphi_{\w}(r^*)$ we obtain
\begin{align}
\label{ScheqS}
    -\partial_{r^*}^2 \varphi_{\w}(r^*) + m^2 g_{00}(r^*) \,  \varphi_{\w}(r^*)=\w^2\varphi_{\w}(r^*).
\end{align}
When $\w \leq m$ the modes decay exponentially at large  $r^* $ and are localized near the horizon \cite{Akhmedov:2016uha}. There is no double degeneration as in  the case of massive field in Rindler space.

The classically permitted region is at the left of the turning point solving the equation $\w^2-m^2 g_{00}(r^*_\text{turning})=0$. Near the horizon where the effective potential vanishes,  the modes approximately behave as
\begin{align}
\label{assymtoticsABC}
     \varphi_\omega(r^*) \approx \sqrt{\frac{2}{\pi}}\cos(\omega r^*+\delta_\omega), \quad \omega < m, \quad \left|\omega r^*\right| \gg 1.
\end{align}
 We do not need their exact form for further considerations.
They do not propagate at spatial infinity and do not contribute to the Hawking radiation \cite{Akhmedov:2015xwa,Akhmedov:2016uha}. On the other hand  they play an important role in the vicinity of the horizon.

When $\w > m$ the situation is similar to the static de Sitter case: there are outgoing modes (right movers) $R_\omega(r^*)$  and ingoing modes (left movers) $L_\omega(r^*)$ which might be represented by resorting to  special functions. We will instead apply  semi-classical approximation methods which in two--dimensions\footnote{Indeed one has that
\begin{align*}
k(x) = \sqrt{(\w r_g)^2 - (m r_g)^2 \, g_{00}(x)}, \ \ \ \ \     \left|\frac{d}{dx} \frac{1}{k(x)}\right| = \frac{1}{2 m r_g} \, \frac{1}{ \big( \frac{\w}{m} - g_{00}(x) \big)^{3/2}} \, \frac{dg_{00}}{dx}.
\end{align*}
Here $ x = \, r_g /r^*$ ;  $k(x)$
is the wave vector of the problem \eqref{ScheqS}; the following inequality holds for all values of
$\omega > m$:
$$
\left|\frac{d}{dx} \frac{1}{k(x)}\right| < 1.
$$
Thus, if $m r_g \gg 1$ the semiclassical approximation is applicable for all values of $r^*$ and there is no  reflection from the potential barrier in \eqref{ScheqS}. Such a reflection is inevitable in four dimensions.  

We would like to thank Dmitriy Trunin for bringing to our attention this important simplification in 2D.}
works well for all values of $r^*$; they lead
to the following solutions:
\begin{align}
\label{WKBR}
    R_\w(r^*)=A_\w \sqrt[4]{\frac{\w^2}{{\w^2-m^2g_{00}
(r^*)}}} \exp\Big(i\ \text{sgn}(\w) \int_{r_0}^{r^*}{\sqrt{\w^2-m^2g_{00}
(x)}}dx\Big),
\\
\label{WKBL}
    L_\w(r^*)=B_\w \sqrt[4]{\frac{\w^2}{{\w^2-m^2g_{00}
(r^*)}}} \exp\Big(-i\ \text{sgn}(\w)\int_{r_0}^{r^*}\sqrt{\w^2-m^2g_{00}
(x)}dx\Big),
\end{align}
where $r_0$ is a reference point.
%
%

Here is the mode expansion of the field operator:
\begin{eqnarray}
\label{fieldopBH}
 &&   \hat{\varphi}(t,r^*)=\int_0^m \frac{d\omega}{\sqrt{2\omega}} \,  e^{-i\omega t} \varphi_\omega(r^*) \hat{c}_\omega+
    \int_m^{+\infty}\frac{d\omega}{\sqrt{2\omega}} e^{-i\omega t}\left[ R_\omega(r^*) \hat{a}_\omega+ L_\omega(r^*) \hat{b}_\omega\right]+h.c.
\end{eqnarray}
where
\begin{eqnarray}
&&    [\hat{a}^{}_{\w},\hat{a}^\dagger_{\w'}]=\delta(\w-\w'), \qquad [\hat{b}^{}_{\w},\hat{b}^\dagger_{\w'}]=\delta(\w-\w'), \qquad [\hat{c}^{}_{\w},\hat{c}^\dagger_{\w'}]=\delta(\w-\w'), \qquad
\end{eqnarray}
all the other commutators being zero.
One may show that in the approximation $m r_g \gg 1$ the canonical commutation relations give
the normalization \begin{align}
\label{ABCWKB}
    |A_\w|^2=|B_\w|^2 \approx \frac{1}{2\pi}.
\end{align}
Taking inspiration from the Rindler and de Sitter cases we may now introduce a Wightman function depending on three (inverse) temperatures as follows: 
%
%
\begin{multline}
\label{WBH}
       W ((t_1,r^*_1) ,\, (t_2, r^*_2))_{\beta_0 \beta_L \beta_R} = \int_{|\omega|<m}\frac{d\w}{2\w} \frac{e^{-i\w(t_1-t_2)}}{1-e^{-\beta_0 \omega}} \, \varphi_\w(r^*_2) \, \varphi_\w(r^*_1) + \\ +
        \int_{|\omega|>m}\frac{d\w}{2\w}\bigg[ \frac{e^{-i\w(t_1-t_2)}}{1-e^{-\beta_L \omega}}L_\w(r^*_1)L^*_\w(r^*_2)+\frac{e^{-i\w(t_1-t_2)}}{1-e^{\beta_R \omega}}R_\w(r^*_1)R^*_\w(r^*_2)\bigg].
\end{multline}
The first term on the RHS of (\ref{WBH}) shows again a double pole at $\omega = 0$. The way we treat it is explained Appendix A.

In Appendix C it is shown  that when $m=0$ and $\beta_{R} = \beta_{L}= 4 \pi r_g$ the above  expression coincides with the  Hartle--Hawking Wightman function; when $\beta_L = \infty$ and $\beta_R = 4 \pi r_g$ it corresponds to the Unruh state; finally,  when $\beta_{R,L} = \infty$ it reproduces the Boulware state.

\subsection{Singularity at the horizon}

The additional singularity at the horizon  is an infrared effect; the main contribution to it comes from the term
%
\begin{equation}
    \varphi_\w(r^*_1)\varphi_\w(r^*_2)\approx\frac{1}{\pi}
    \cos\Big(\omega(r^*_1+r^*_2)+2\delta_\omega\Big)
\end{equation}
 in \eqref{WBH},  with $\w \to 0$.
To fix the phase we consider space--like separated points near the horizon  parametrized as follows:
\begin{align}\label{horlimit}
    r^*_1=\lambda, \quad r^*_2=\lambda + const, \quad t_1=t_2 = - \lambda.
\end{align}
The (future) horizon limit corresponds to $\lambda\to-\infty$; in this limit the Wightman function has the following asymptotics:
\begin{align}
\label{wm}
    W(t_1,r^*_1=\lambda | t_2, r^*_2=\lambda)=\frac{\lambda}{\beta_0}e^{2i\delta_0}, \qquad\text{as} \ \ \lambda\to-\infty.
\end{align}
As anticipated the limiting expression depends on the phase and goes to zero  in the zero temperature limit.
At low energies turning point
\begin{align}
 r^*_\text{turning} 
 \approx r_g \log\frac{\w^2}{m^2},\qquad\text{as} \ \ \w\to0.
\end{align}
is shifted to minus infinity. Since in this limit  the  Rindler space asymptotics should be reproduced we have to set $\delta_0=\frac{\pi}{2}$.
In Appendix D we present another derivation of  this equality.

\subsection{Stress-energy tensor}

In the lightcone coordinates
    $V = t+r^*, \ \  U = t-r^*$
the metric \eqref{BHmetr} takes the form
\begin{align}
    ds^2 = C(U,V) dU dV, \ \ \ \
    C(U,V) = \frac{{\mathcal W}(e^{\frac{V-U}{4M}-1})}{1 + {\mathcal W}(e^{\frac{V-U}{4M}-1})},
\end{align}
where  ${\mathcal W}(r^*)$ is the Lambert function.  
Near the horizon
\begin{multline}
    W((V^+,U^+),(V^-,U^-)) \approx \int^m_{-m} \frac{d \w }{4 \pi \w} \frac{1}{e^{\beta_0 \w}-1} \bigg( e^{i \w (V^+ - U^-)+2i \delta_\w }+ e^{i \w (V^+-V^-)} + e^{i \w (U^+-U^-)}+\\
    +e^{i\w (U^+-V^-)-2i \delta_\w} \bigg)
    + \int_{|\w|>m} \frac{d \w }{4 \pi \w} \bigg[ \frac{e^{i\w (V^+-V^-) }}{e^{\beta_R \w}-1} + \frac{e^{i\w (U^+-U^-) }}{e^{\beta_L \w}-1} \bigg],
\end{multline}
 By taking the limit of coinciding points one gets
\begin{multline}
    \partial_{U^+} \partial_{U^-} W  \approx - \frac{1}{4\pi} \frac{1}{(U^+-U^-)^2} +  \frac{\pi}{12 \beta_0^2} + \frac{1}{2\pi} \bigg( \frac{Li_2 (e^{-m \beta_L)}}{\beta_L^2} - \frac{Li_2 (e^{-m \beta_0)}}{\beta_0^2} \bigg) + \\
    +\frac{m}{2\pi} \bigg( \frac{\log (1-e^{-m \beta_0} )}{\beta_0} -  \frac{\log (1-e^{-m \beta_L} )}{\beta_L} \bigg),
\end{multline}
where $Li_2(x)$ is the polylogarithmic function.

Then for the components of the stress-energy tensor we obtain
\begin{multline}
    T_{UU} \approx - \bigg[ \frac{1}{4 \pi \epsilon^2 (t_{\alpha} t^{\alpha} )} + \frac{R}{24 \pi} \bigg] \frac{t_U t_U}{t_{\alpha} t^{\alpha}} + \frac{\pi}{12} \bigg( \frac{1}{\beta_0^2} - \frac{1}{(8 \pi M)^2} \bigg) + \frac{1}{2\pi} \bigg( \frac{Li_2 (e^{-m \beta_L)}}{\beta_L^2} - \frac{Li_2 (e^{-m \beta_0)}}{\beta_0^2} \bigg) +\\
    +\frac{m}{2\pi} \bigg( \frac{\log (1-e^{-m \beta_0} )}{\beta_0} -  \frac{\log (1-e^{-m \beta_L} )}{\beta_L} \bigg).
\\
    T_{VV} \approx - \bigg[ \frac{1}{4 \pi \epsilon^2 (t_{\alpha} t^{\alpha} )} + \frac{R}{24 \pi} \bigg] \frac{t_V t_V}{t_{\alpha} t^{\alpha}} + \frac{\pi}{12} \bigg( \frac{1}{\beta_0^2} - \frac{1}{(8 \pi M)^2} \bigg) + \frac{1}{2\pi} \bigg( \frac{Li_2 (e^{-m \beta_R)}}{\beta_R^2} - \frac{Li_2 (e^{-m \beta_0)}}{\beta_0^2} \bigg) + \\
    +\frac{m}{2\pi} \bigg( \frac{\log (1-e^{-m \beta_0} )}{\beta_0} -  \frac{\log (1-e^{-m \beta_R} )}{\beta_R} \bigg).
\end{multline}
The non-diagonal component near the horizon goes to zero
\begin{align} \label{nondiag_schw}
    T_{UV} \approx \frac{m^2}{4} e^{{\lambda}/{2 r_g}} \frac{| \lambda|}{\beta_0} \to 0.
\end{align}
Near the horizon the stress--energy tensor in similar to the original result of \cite{Davies:1976ei,Davies:1977}
\begin{align} \label{set_schwar}
    T_{\mu \nu} \approx \Theta_{\mu \nu} + \frac{R}{48 \pi } g_{\mu \nu},
\end{align}
where
\begin{align*}
    \Theta_{UU} &= -\frac{1}{12 \pi} C^{1/2} \partial^2_U C^{-1/2}+ \frac{\pi}{12 \beta_0^2}  + L(\beta_L, \beta_0),\\
    \Theta_{VV} &= -\frac{1}{12 \pi} C^{1/2} \partial^2_V C^{-1/2} + \frac{\pi}{12 \beta_0^2} + L(\beta_R,\beta_0),\\
    \Theta_{UV} &= \Theta_{VU} =0,
\end{align*}
and
\begin{align}
    L(\beta_1, \beta_2) = \bigg( \frac{Li_2 (e^{-m \beta_1})}{\beta_1^2} - \frac{Li_2 (e^{-m \beta_2})}{\beta_2^2} \bigg)
    +\frac{m}{2\pi} \bigg( \frac{\log (1-e^{-m \beta_2} )}{\beta_2} -  \frac{\log (1-e^{-m \beta_1} )}{\beta_1} \bigg).
\end{align}
Some comments are in order here. When the three temperatures coincide the finite logarithmic and dilogarithmic contributions vanish. Furthermore, there are no finite contributions at all when they all are equal to the Hawking temperature $\beta = 4 \pi r_g$. Second, while the additional singularity of the propagators is effective only on the non-diagonal components of stress-energy tensor (in $(u,v)$ coordinates), the exponential damping protects the covariant components \eqref{nondiag_schw}. Additional singularity does arise in  the  mixed components of stress energy tensor as follows:
\begin{align}
    T^{V}_{\ \ V } =  T^{U}_{\ \ U} =\frac{m^2}{2}  \langle \varphi \varphi \rangle \sim \frac{m^2 }{2 \beta_0} | \lambda |, \qquad\text{as} \ \ \lambda\to-\infty.
\end{align}

\section{Outlook}

Heating and thermalization are however non--stationary processes.
To calculate the correlation functions in non--stationary situations
one has to exploit the Schwinger--Keldysh diagrammatic technique \cite{LL10,Kamenev}. The starting point is to choose an initial Cauchy surface\footnote{This method applies to globally hyperbolic spacetimes. In non globally hyperbolic space--times  one should deal also with boundary conditions \cite{Akhmedov:2018lkp}.} and an initial value of the correlation functions, i.e. an initial state. The Schwinger--Keldysh technique provides the time evolution towards future of the correlation function in question.

Different types of Cauchy surfaces and initial values may and in general do lead to substantially different physical behaviours 
\cite{Akhmedov:2013vka,Akhmedov:2019cfd}. 
Even in highly symmetric curved space--time (such as de Sitter)  the tree--level correlators of a generic state are not functions of geodesic distances. It goes without saying about generic space--times, which only partly resemble to the de Sitter space \cite{Akhmedov:2019cfd}. In this sense the situation in strongly curved space--times is similar to the condensed matter phenomena rather than to high energy physics ones.

There is no a priori reason for the initial state in the early universe or in the vicinity of primordial black holes be necessarily the ground state or a thermal state at the canonical temperature. Here we considered a class of time translation invariant states in Rindler, static de Sitter and two--dimensional black hole space--times. They can be thought as initial states for thermalization or heating problems. They may also appear as attractor equilibrium states at the end  of some process.  We have shown that when the various temperatures  do not coincide with the canonical ones then the two--point Wightman functions have anomalous singularities at the horizons. That may affect the loop corrections. The latter are necessary to calculate and to resum to trace the fate of the initial state and of the correlation functions. See e.g.  \cite{LL10,Kamenev,Akhmedov:2013vka} for various related situations and \cite{Mirbabayi:2020vyt} for the recent study of the resummation of loop corrections in static patch for a particular initial state. Loop corrections for various initial states in static patch will be considered elsewhere.

\section{Acknowledgements}

We would like to acknowledge valuable discussions with O.Diatlyk,  F. Popov, A.Semenov and D.Trunin.

The work of ETA was supported by the grant from the Foundation for the Advancement of Theoretical Physics and Mathematics ``BASIS'' and by RFBR grant 18-01-00460. The work of ETA, PAA, KVB and DVD is supported by Russian Ministry of education and science (project 5-100).


\newpage
\begin{appendices}

\numberwithin{equation}{section}

\setcounter{equation}{0}
\renewcommand\theequation{A.\arabic{equation}}

\section{Leading infrared contribution}
\label{appendixA}
 The behavior of various Wightman functions discussed above at the horizons is governed by the integral of the form:
\begin{align*}
\int_{-\infty}^{+\infty} \frac{d \omega}{\omega+i\epsilon}\frac{e^{i \omega \theta}}{e^{\beta(\omega+i\epsilon)}-1}, \quad \text{where} \ \ |\theta|\gg 1.
\end{align*}
 The choice of the shifts of the poles here  reproduces the results in the case  $\beta = 2\pi/N$ but it can also justified by general distributional methods \cite{gelfand}. 
The contour in closed in the upper half-plane for positive values of $\theta$ and in the lower half for negative ones. In the first case the double pole at $\omega=-i\epsilon$ does not contribute.  Contributions from other poles are suppressed. 
For negative $\theta$ the leading contributions in the limit $\theta \to - \infty$ comes from the double pole at $\omega=-i\epsilon$:
\begin{align}
\label{helpfull}
    \int_{-\infty}^{+\infty} \frac{d\omega}{\omega+i\epsilon}\frac{e^{i\omega\theta}}{e^{\beta(\omega+i\epsilon)}-1}\approx\begin{cases}
   0 \quad \text{if} \ \ \theta>0\\
   \frac{2\pi}{\beta} \theta \quad \text{if} \ \ \theta<0
 \end{cases}
 \quad \text{as} \ \ |\theta|\gg 1.
\end{align}
the answer depends on the sign of $\theta$.


\section{Point-splitting regularization}
\renewcommand\theequation{B.\arabic{equation}}
\setcounter{equation}{0}
\label{appendixD}

\begin{figure}[h]
    \centering
    \begin{tikzpicture}[
    scale=2]
  \coordinate (x) at (0, 0);
  \coordinate (xp) at (1.5, 0.75);
  \coordinate (xpp) at (3, 0.5);
  \coordinate (xm) at (-1.5, -0.5);
  \coordinate (xmm) at (-3, 0.5);
  \coordinate (xt) at (0.5,0.86);
  \draw[thick] (xmm) to [out=20,in=150] (xm) node[circle,fill,inner sep=1.5pt]{} to [out=-30,in=-120] (x) node[circle,fill,inner sep=1.5pt]{} to [out=60,in=140] (xp) node[circle,fill,inner sep=1.5pt]{} to [out=-40,in=-170] (xpp);

  \node at (x) [left]{\LARGE $x$ \normalsize};
  \node at (1.75,0.75) [above]{\LARGE $x^+$ \normalsize};
  \node at (xm) [below]{\LARGE $x^-$ \normalsize};

   \draw[thick,->] (x) -- (xt);

   \node at (xt) [above]{\LARGE $t^\mu$ \normalsize};

   \draw [thick, ->] (-3,-1) -- node[midway, right] {geodesic line} (-2.75,0.25);
\end{tikzpicture}
    \caption{Point-splitting}
    \label{geopic}
\end{figure}
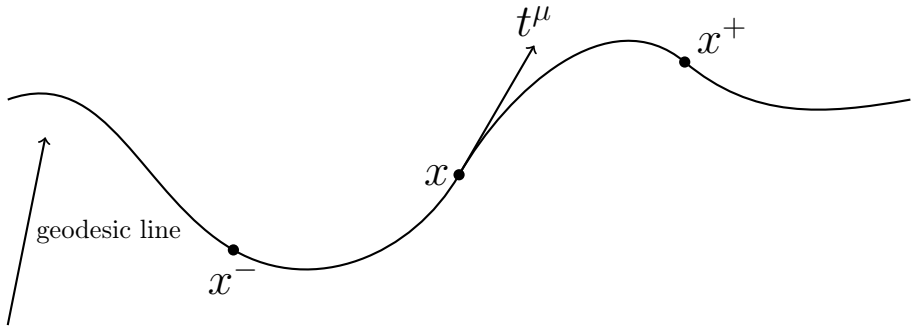

To make the paper self-contained and set up the notations in this appendix we summarize here the standard point splitting regularization procedure \cite{Davies:1976ei} of the  expectation value of the stress--energy tensor in curved space--time:
\begin{align*}
    \langle \hat{T}_{\mu\nu}(x) \rangle = \left. D_{\mu\nu} \langle \hat{\varphi}(x^+)\hat{\varphi}(x^-) \rangle\right|_{x^+=x^-=x}.
\end{align*}
Here $D_{\mu\nu}$ is a differential operator; $x^\pm$ are points which are separated from $x$ along a  \textcolor{black}{spacelike} geodesic  and $t^\mu$ is tangent  vector (see the Fig. \ref{geopic}).
A point  close enough to $x^\mu$ can be represented as follows
\begin{align}
\label{eqgeo}
    x^\mu(\tau)=x^\mu+\tau t^\mu+\frac{1}{2} \tau^2 a^\mu+\frac{1}{6}\tau^3 b^\mu+...,
\end{align}
where $\tau$ is the proper length. 
And the coordinates of $x^\pm$ are
$x^{\mu}{}^\pm=x^{\mu}(\tau=\pm \epsilon).$

General two dimensional conformally flat metric can be written $ds^2=C(u,v)dudv$.
%
The geodesic equations provide the relations between the parameters
$t^\mu, a^\mu, b^\mu$: 
\begin{align}
    a^\mu=-\Gamma^\mu_{\nu\lambda}t^\nu t^\lambda, \quad b^\mu=-\Gamma^\mu_{\nu\lambda}(a^\nu t^\lambda+t^\nu a^\lambda)-t^\sigma \partial_\sigma \Gamma^\mu_{\nu\lambda}t^\nu t^\lambda.
\end{align}
It is enough to express $a^\mu$ and $b^\mu$ in terms of $t^\mu$ to find the finite part of the expectation value of the stress--energy tensor. Another building block 
is the parallel transport matrix $e^\mu_\nu(\tau)$, solving the following equation:
\begin{align}
    \frac{d e^\mu_\nu}{d\tau}+\Gamma^\mu_{\rho \sigma} \frac{dx^\rho}{d\tau}e^{\sigma}_{\nu} =0, \ \ 
    e^{\mu}_\nu(\tau=0)=\delta_\nu^\mu.
\end{align}
Again, one expands the  parallel transport matrix in powers of $\tau$:
\begin{align*}
    e^\mu_\nu=\delta^\mu_\nu+\tau t^\mu_\nu+\frac{1}{2}\tau^2 a^\mu_\nu+...
\end{align*}
where
\begin{align*}
    t^\mu_\nu=-\Gamma^\mu_{\rho \nu}t^\rho, \quad a^\mu_\nu=\Gamma^\mu_{\rho \nu}\Gamma^\rho_{\alpha\beta}t^\alpha t^\beta+\Gamma^\mu_{\rho\sigma}\Gamma^\sigma_{\alpha\nu}t^\rho t^\alpha-t^\alpha t^\rho \partial_\alpha \Gamma^\mu_{\rho \nu}.
\end{align*}
The expectation value of the stress--energy tensor in a state at inverse  temperature $\beta^{-1}$ is given by
\begin{align}
\label{genT}
    \langle \hat{T}_{\mu\nu} \rangle_\beta=\langle \partial_\alpha \varphi(x^+)\partial_\beta \varphi(x^-)\rangle_\beta\left( e^{+\alpha}_\mu e^{-\beta}_\nu - \frac{1}{2}g_{\mu\nu} g^{\sigma\rho}e^{+\alpha}_\sigma e^{-\beta}_\rho\right)+\frac{1}{2}m^2g_{\mu\nu} \langle\varphi(x^+)\varphi(x^-)\rangle_\beta.
\end{align}
where $e_\mu^{\pm \alpha}=e_\mu^\alpha(\tau=\pm\epsilon)$ 
and  the limit $\epsilon\to 0$ is taken. 
The result will contain terms that depend on $\epsilon$ and direction-dependent terms. For example, for the massless field in the generic conformally flat background one has \cite{Davies:1977}:
\begin{align}
    \langle   T_{\mu \nu}\rangle = - \bigg[ \frac{1}{4 \pi \epsilon^2 (t_{\alpha} t^{\alpha})} + \frac{R}{24 \pi} \bigg] \bigg[ \frac{t_{\mu} t_{\nu}}{t_{\alpha} t^{\alpha}} - \frac{1}{2} g_{\mu \nu} \bigg] + \Theta_{\mu \nu},
\end{align}
and the regularized stress--energy tensor reads:
\begin{align}
    \langle   :T_{\mu \nu}: \rangle =\Theta_{\mu \nu} + \frac{R }{48 \pi } g_{\mu \nu},
\end{align}
with
\begin{align}
    \Theta_{uu} &= -\frac{1}{12 \pi} C^{1/2} \partial^2_u C^{-1/2}+ \text{state dependent terms},\\
    \Theta_{vv} &= -\frac{1}{12 \pi} C^{1/2} \partial^2_v C^{-1/2} + \text{state dependent terms},\\
    \Theta_{uv} &= \Theta_{vu} =0.
\end{align}
The tensor is conserved for an invariant state 
%
 only if one omits the direction--dependent terms. While averaging over directions leads to quantities, which are not covariantly conserved.


\section{Boulware, Unruh and Hartle--Hawking states}
\label{appendixB}
\setcounter{equation}{0}
\renewcommand\theequation{C.\arabic{equation}}

There are several different ways to define Boulware, Unruh and Hartle--Hawking states for {\it massless scalar fields in four--dimensions}. Not all of them can be straightforwardly generalized to the  massive case.
 Here we repeat the standard constructions and  consider their generalizations to the massive case.


\subsection{Analytic continuation of the positive frequency modes}



We look for a  complete set of solutions of the massless  Klein-Gordon equation in either the left or right (Schwarzschild) quadrant of the entire black--hole space--time in four dimensions (see fig. \ref{Schpic}). We require these functions to have definite sign of frequency with respect to the time-like Killing vector $\frac{\partial}{\partial t}$ (in the left quadrant $ - \frac{\partial}{\partial t}$):

\begin{eqnarray} \label{set1}
&\overrightarrow{u}_{\omega lm} (x) = (4 \pi \omega)^{-1/2} e^{-i \omega t} \overrightarrow{R}_l(\omega | r) Y_{lm} (\theta, \varphi), \nonumber \\
&\overleftarrow{u}_{\omega lm} (x) = (4 \pi \omega)^{-1/2} e^{-i \omega t} \overleftarrow{R}_l(\omega | r) Y_{lm} (\theta, \varphi),
\end{eqnarray}
where $\overrightarrow{R}_l(\omega | r)$ and $\overleftarrow{R}_l(\omega | r)$ are solutions of the radial equation  corresponding to outgoing and incoming waves  respectively \cite{Candelas:1980zt} and  $Y_{lm} (\theta, \varphi)$ are the standard spherical harmonics;
$x = (t,r,\theta, \varphi)$. The Boulware two-point function is 
\begin{equation} \label{boulware4d}
W_B (x,x')= 
  \sum_{lm} \int^{\infty}_0 \frac{d \omega}{4 \pi \omega} e^{-i \omega (t-t')} Y_{lm}(\theta,\phi) Y^*_{lm}(\theta',\phi') \left[ \overrightarrow{R}_l (\omega|r) \overrightarrow{R}^*_l (\omega|r')+\overleftarrow{R}_l (\omega|r) \overleftarrow{R}^*_l (\omega|r') \right].
\end{equation}
To define the Unruh state 
 the Kruskal extension is needed: 
%
\begin{eqnarray} \label{kruskal}
U=t-r-2M \log(r/2M-1), \ \ \Tilde{U}=-4Me^{-U/4M}, \nonumber \\
V=t+r+2M \log(r/2M-1), \ \ \ \ \ \ \ \Tilde{V}=4M e^{V/4M}. \nonumber
\end{eqnarray}
The Unruh modes are  positive--frequency w.r.t. $U$ and near the paste horizon  behave as follows:
\begin{equation}
y_{\omega lm} \sim e^{-i \omega \Tilde{U}} Y_{lm}(\theta, \varphi).\label{set2}
\end{equation}
They are analytic functions in the lower half-plane of the complex variable $\Tilde{U}$. 
On the other hand the behaviour of the modes (\ref{set1}), 
on the  past horizon of the right patch and, respectively, on the  future horizon of left patch, is as follows
\begin{eqnarray}
&&\overrightarrow{u}_{\omega lm}^R (x) \approx (4 \pi \omega)^{-1/2} \left|\frac{\Tilde{U}}{4M}\right|^{i4M\omega} Y_{lm} (\theta, \varphi),
\\
&&    \overrightarrow{u}_{\omega lm}^L (x) \approx (4 \pi \omega)^{-1/2} \left|\frac{\Tilde{U}}{4M}\right|^{-i4M\omega} Y_{lm} (\theta, \varphi).
\end{eqnarray}
Then the normalized combinations
\begin{equation}
\overrightarrow{y}_{\omega lm} = \frac{1}{\sqrt{|2 \sinh (4 \pi M \omega)}|} \Big[e^{2 \pi M \omega} \overrightarrow{u}^R_{\omega lm} + e^{-2 \pi M \omega} (\overrightarrow{u}^{L}_{\omega lm})^*\Big],
\end{equation}
have the same analyticity properties  of the modes (\ref{set2}) and are equivalent to them.
We can then compute Wightman function of Unruh state when the two points are located in the right Schwarzschild patch:
\begin{equation} \label{unruh4d}
W_U (x,x') 
 = \sum_{lm} \int^{+\infty}_{-\infty} d \omega \bigg[  \frac{\overrightarrow{u}_{\omega lm}(x) \overrightarrow{u}^*_{\omega lm}(x')}{1-e^{-\frac{2 \pi \omega}{\kappa}}}
 +\overleftarrow{u}_{\omega lm}(x) \overleftarrow{u}^*_{\omega lm}(x') \theta(\omega) \bigg],
\end{equation}
where $\kappa = (4M)^{-1}$ is the surface gravity.
In a similar manner, 
the   modes that are positive frequency  w.r.t. $ \frac{\partial}{\partial \Tilde{V}}$   are:
\begin{equation}
\overleftarrow{y}_{\omega lm} = \frac{1}{\sqrt{2 \sinh (4 \pi M \omega)}} \Big[e^{-2 \pi M \omega} (\overleftarrow{u}^R_{\omega lm})^* + e^{2 \pi M \omega} \overleftarrow{u}^L_{\omega lm}\Big].
\end{equation}
They give rise to  the  Hartle-Hawking  Wightman function:
\begin{equation} \label{hh4d}
W_H (x,x') 
=  \sum_{lm} \int^{+\infty}_{-\infty} {d \omega} \bigg[  \frac{\overrightarrow{u}_{\omega lm}(x) \overrightarrow{u}^*_{\omega lm}(x')}{1-e^{-\frac{2 \pi \omega}{\kappa}}}
 +
  \frac{\overleftarrow{u}^*_{\omega lm}(x) \overleftarrow{u}_{\omega lm}(x')}{e^{\frac{2 \pi \omega}{\kappa}}-1} \bigg]
\end{equation}
The outgoing waves of the Unruh state are thermally distributed  at temperature $T=\frac{\kappa}{2 \pi} = \frac{1}{8 \pi M}$;   both the outgoing and incoming waves, of the Hartle-Hawking state  are thermally distributed  at the same temperature. In the two dimensional case these formulae reduce to:
\begin{eqnarray}
    W_U (x,x')
 = \int^{+\infty}_{-\infty} d \omega \bigg[  \frac{\overrightarrow{u}_{\omega}(x) \overrightarrow{u}^*_{\omega}(x')}{1-e^{-\frac{2 \pi \omega}{\kappa}}}
 +\overleftarrow{u}_{\omega}(x) \overleftarrow{u}^*_{\omega}(x') \theta(\omega) \bigg], \nonumber \\
    W_H (x,x')
=  \int^{+\infty}_{-\infty} {d \omega} \bigg[  \frac{\overrightarrow{u}_{\omega}(x) \overrightarrow{u}^*_{\omega}(x')}{1-e^{-\frac{2 \pi \omega}{\kappa}}}
 +
  \frac{\overleftarrow{u}^*_{\omega}(x) \overleftarrow{u}_{\omega}(x')}{e^{\frac{2 \pi \omega}{\kappa}}-1} \bigg]. \label{WUWH1}
\end{eqnarray}

\subsection{Two dimensions again}
The following construction  is valid for any stationary background, provided there are left and right movers. Stationarity implies that a  Wightman function with zero anomalous quantum averages depends only on the difference of times:

\begin{multline}
    W (x,x')
= \int^{+\infty}_{0} {d \omega} \int^{+\infty}_{0} {d \omega'} \bigg[ \langle a_{\omega} a_{\omega'}^{\dagger} \rangle \overrightarrow{u}_{\omega}(x) \overrightarrow{u}^*_{\omega'}(x')
 + \langle a_{\omega}^{\dagger} a_{\omega'} \rangle \overrightarrow{u}^*_{\omega}(x) \overrightarrow{u}_{\omega'}(x')
 + \\
 +\langle b_{\omega} b_{\omega'}^{\dagger} \rangle\overleftarrow{u}_{\omega}(x) \overleftarrow{u}^*_{\omega'}(x') + \langle b_{\omega}^{\dagger} b_{\omega'} \rangle
  \overleftarrow{u}^*_{\omega}(x) \overleftarrow{u}_{\omega'}(x') \bigg].
\end{multline}
In this general setting the Unruh and the Hartle-Hawking states correspond respectively to the following choices
\begin{equation}
     \langle a_{\omega}^{\dagger} a_{\omega'} \rangle = \frac{1}{e^{\frac{2 \pi \omega}{\kappa}}-1} \delta(\w -\w'), \quad \langle b_{\omega}^{\dagger} b_{\omega'} \rangle = 0
\end{equation}
\begin{equation}
     \langle a_{\omega}^{\dagger} a_{\omega'} \rangle = \langle b_{\omega}^{\dagger} b_{\omega'} \rangle = \frac{1}{e^{\frac{2 \pi \omega}{\kappa}}-1}\delta(\w -\w'),
\end{equation}
so that
\begin{eqnarray} \label{WUWH}
   && W_U (x,x')
 = \int^{+\infty}_{0} d \omega \bigg[  \frac{\overrightarrow{u}_{\omega}(x) \overrightarrow{u}^*_{\omega}(x')}{1-e^{-\frac{2 \pi \omega}{\kappa}}}
 + \frac{\overrightarrow{u}^*_{\omega}(x) \overrightarrow{u}_{\omega}(x')}{e^{\frac{2 \pi \omega}{\kappa}}-1}
 +\overleftarrow{u}_{\omega}(x) \overleftarrow{u}^*_{\omega}(x')  \bigg] , \\
&& W_H (x,x')
= \int^{+\infty}_{0} {d \omega} \bigg[  \frac{\overrightarrow{u}_{\omega}(x) \overrightarrow{u}^*_{\omega}(x')}{1-e^{-\frac{2 \pi \omega}{\kappa}}}
 + \frac{\overrightarrow{u}^*_{\omega}(x) \overrightarrow{u}_{\omega}(x')}{e^{\frac{2 \pi \omega}{\kappa}}-1}
 +\frac{\overleftarrow{u}_{\omega}(x) \overleftarrow{u}^*_{\omega}(x')}{1-e^{-\frac{2 \pi \omega}{\kappa}}} +
  \frac{\overleftarrow{u}^*_{\omega}(x) \overleftarrow{u}_{\omega}(x')}{e^{\frac{2 \pi \omega}{\kappa}}-1} \bigg] \cr &&
\end{eqnarray}
These expressions are equivalent to (\ref{WUWH1}) if the following condition holds:
\begin{align} \label{condition}
    u_{-\w}^* (x) u_{-\w} (x') = - u_\w (x) u_\w^*(x')
\end{align}
for both outgoing and incoming waves; 
this condition is verified in the two-dimensional Schwarzschild spacetime. 

In  the massive  case, gluing modes  \eqref{assymtoticsABC} in the classically permitted and forbidden regions gives the relation
\begin{equation}
    e^{2 i \delta_{\w}} = \frac{i \omega - \sqrt{m^2-\w^2}}{i \w + \sqrt{m^2-\w^2}} e^{-2 i \w r^*_\text{turning}}.
\end{equation}
From here one can show that
\begin{align*}
    \varphi_\w(r^*_1)\varphi_\w(r^*_2) = \varphi_{-\w}(r^*_1)\varphi_{-\w}(r^*_2),
\end{align*}
and this provides another justification of Eq.  \eqref{WBH}. 

\label{appendixC}
\setcounter{equation}{0}
\renewcommand\theequation{D.\arabic{equation}}

In the horizon limit 
modes as with  $|\w| < m$ behave as follows
\begin{equation}
    \phi_\w (r^*) =  C(\w) K_{4 i \w M} (4M m \xi) \sim \cos(\w r^* + \delta_\w), \ \ \  \xi^2 = \frac{r}{2M}-1
\end{equation}
with
\begin{equation}
    \delta_\w = \frac{\pi}{2} + r_g \w \big( 2 \log(m r_g)-1 \big) - {\rm arg}\, \Gamma (1+ i \w r_g).
\end{equation}
It follows that $\delta_0 = \frac{\pi}{2} $. Noting that ${\rm arg}\, \Gamma (1+ i \w r_g) = - {\rm arg}\, \Gamma (1- i \w r_g)$  again points towards \eqref{WBH}.

\end{appendices}

\end{document}